\documentclass[pre,twocolumn,showpacs,floatfix]{revtex4}
 \usepackage{graphicx}
 \usepackage{epstopdf}
 \usepackage{dcolumn}
 \usepackage{color}
 \usepackage{bm}
 \usepackage{verbatim}
 \usepackage{multirow}
 \usepackage{amsmath,amssymb,graphicx}
 \usepackage{epsfig}
 \usepackage{enumerate}
 \usepackage{array}
 \usepackage{multirow}
\newcommand{\figOne}
{
 \begin{figure*}[t]
 \begin{center}
 \includegraphics[width=2.5in,height=1.75in]{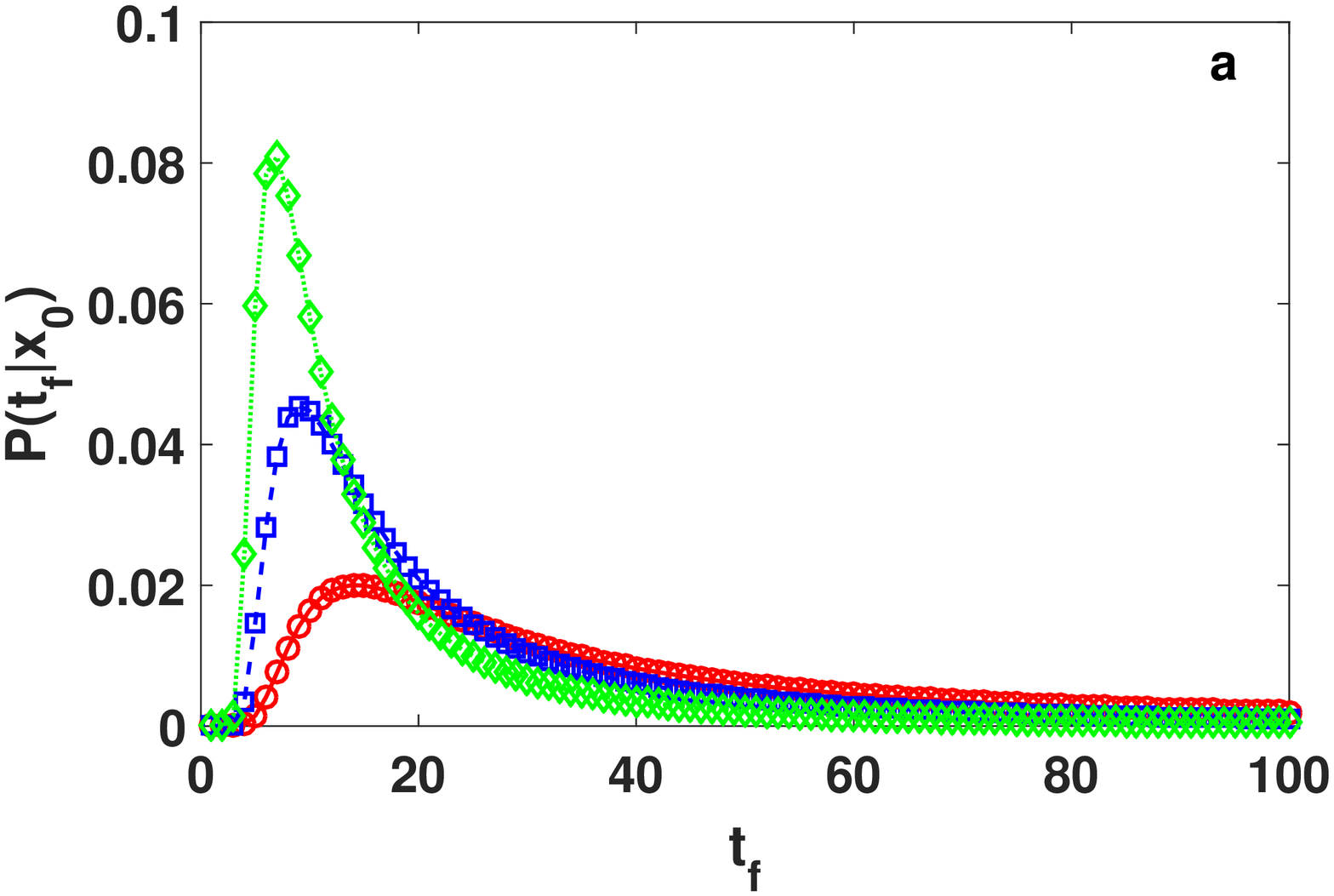}
 \includegraphics[width=2.5in,height=1.75in]{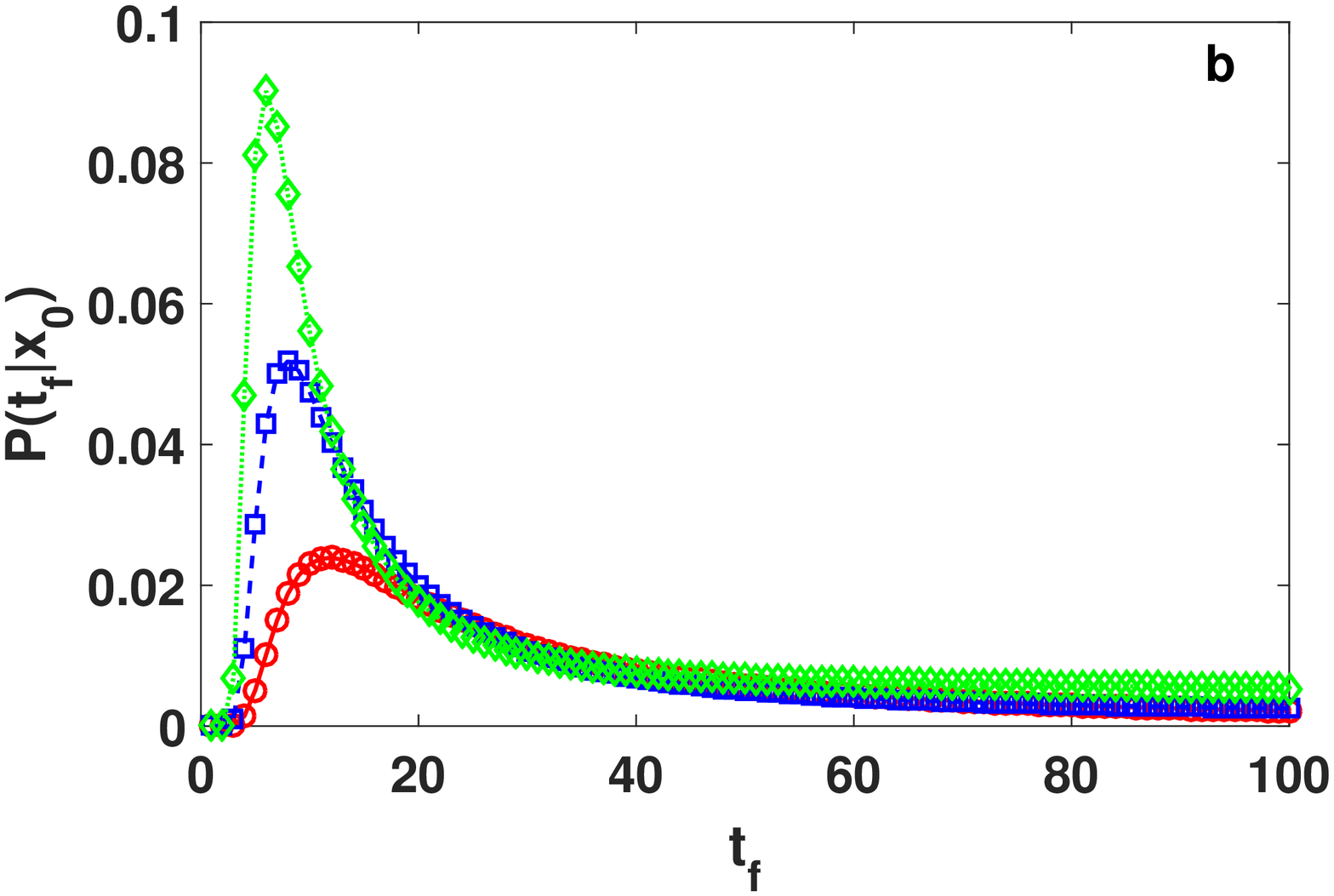}
 \caption{(color online) Plot of first passage time densities for the (a) unbiased case,i.e., $\mu(t)=0$ and $D(t)=2kt^{\alpha}$  (b)biased case with proportional power-law diffusion and drift, i.e., $\mu(t)=qkt^{\alpha}$ and $D(t)=2k t^{\alpha}$. Here, we use $q=0.1$, $k=1.0$, $\alpha=0.5$ (red circle); $\alpha=1.0$ (blue square) and $\alpha=1.5$ (Green triangle).}
\label{fig:energy}
\end{center}
\end{figure*}
}

\newcommand{\figThree}
{
 \begin{figure*}[t]
 \begin{center}
 \includegraphics[width=2.5in,height=1.75in]{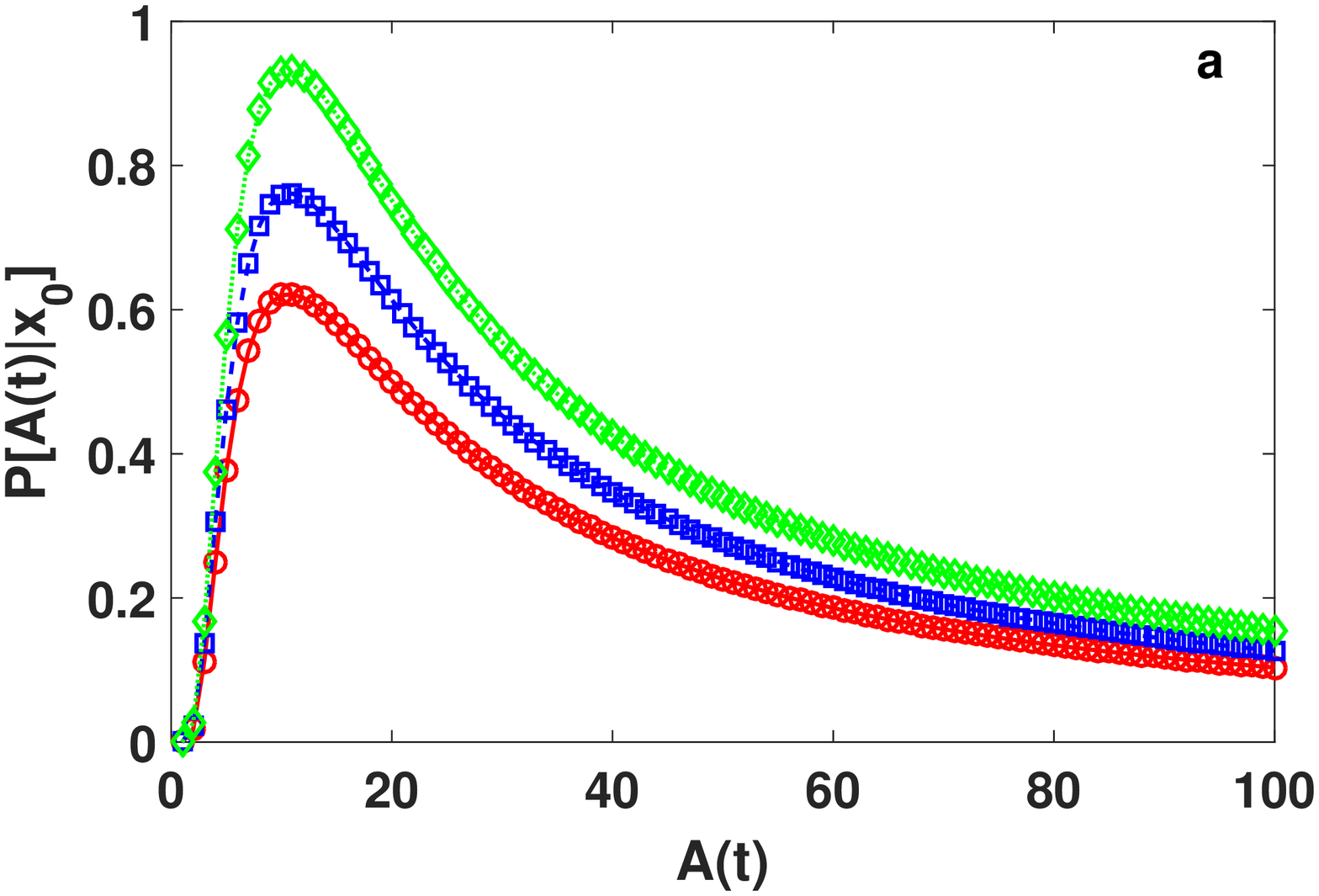}
 \includegraphics[width=2.5in,height=1.75in]{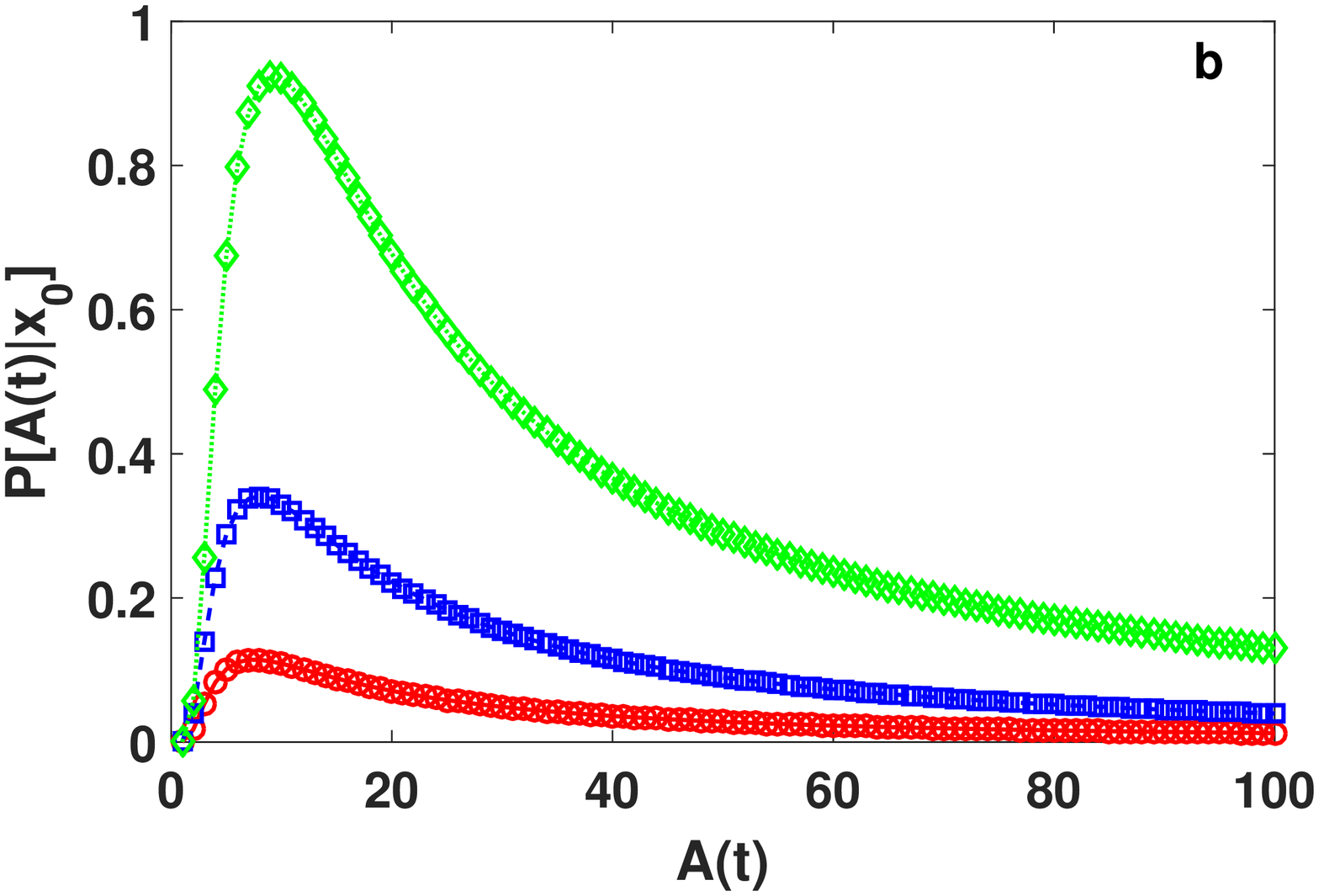}
 \caption{(color online) Plot of pdf of area A(t) before first passage time  for the (a) unbiased case,i.e., $\mu(t)=0$ and $D(t)=2kt^{\alpha}$  (b)biased case with proportional power-law diffusion and drift, i.e., $\mu(t)=qkt^{\alpha}$ and $D(t)=2k t^{\alpha}$. Here, we use $q=0.1$, $k=1.0$, $\alpha=0.5$ (red circle); $\alpha=1.0$ (blue square) and $\alpha=1.5$ (Green triangle).}
\label{fig:energy}
\end{center}
\end{figure*}
}

\newcommand{\figFive}
{
 \begin{figure*}[h]
 \begin{center}
 \includegraphics[width=3in,height=2.3in]{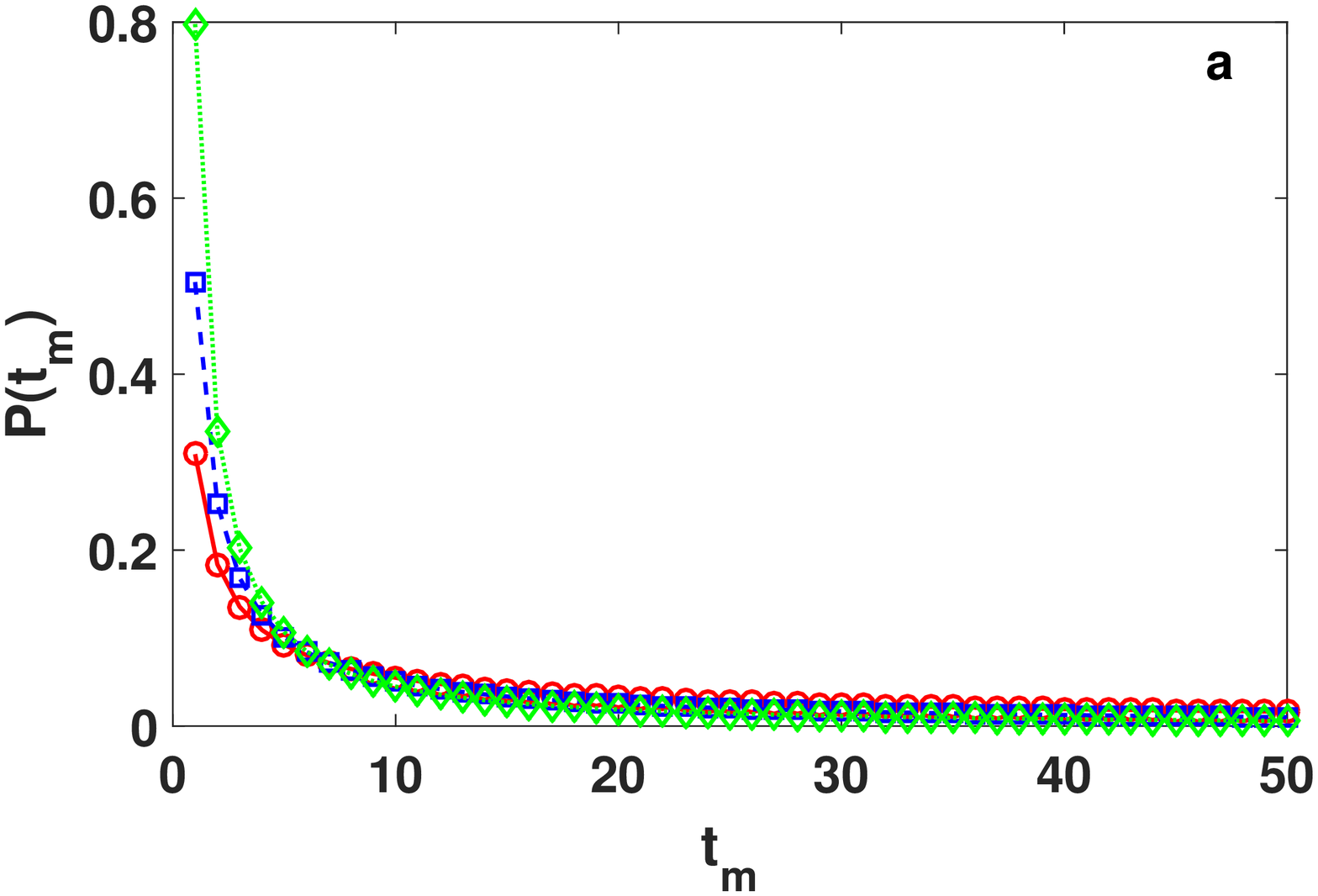}
 \includegraphics[width=3in,height=2.3in]{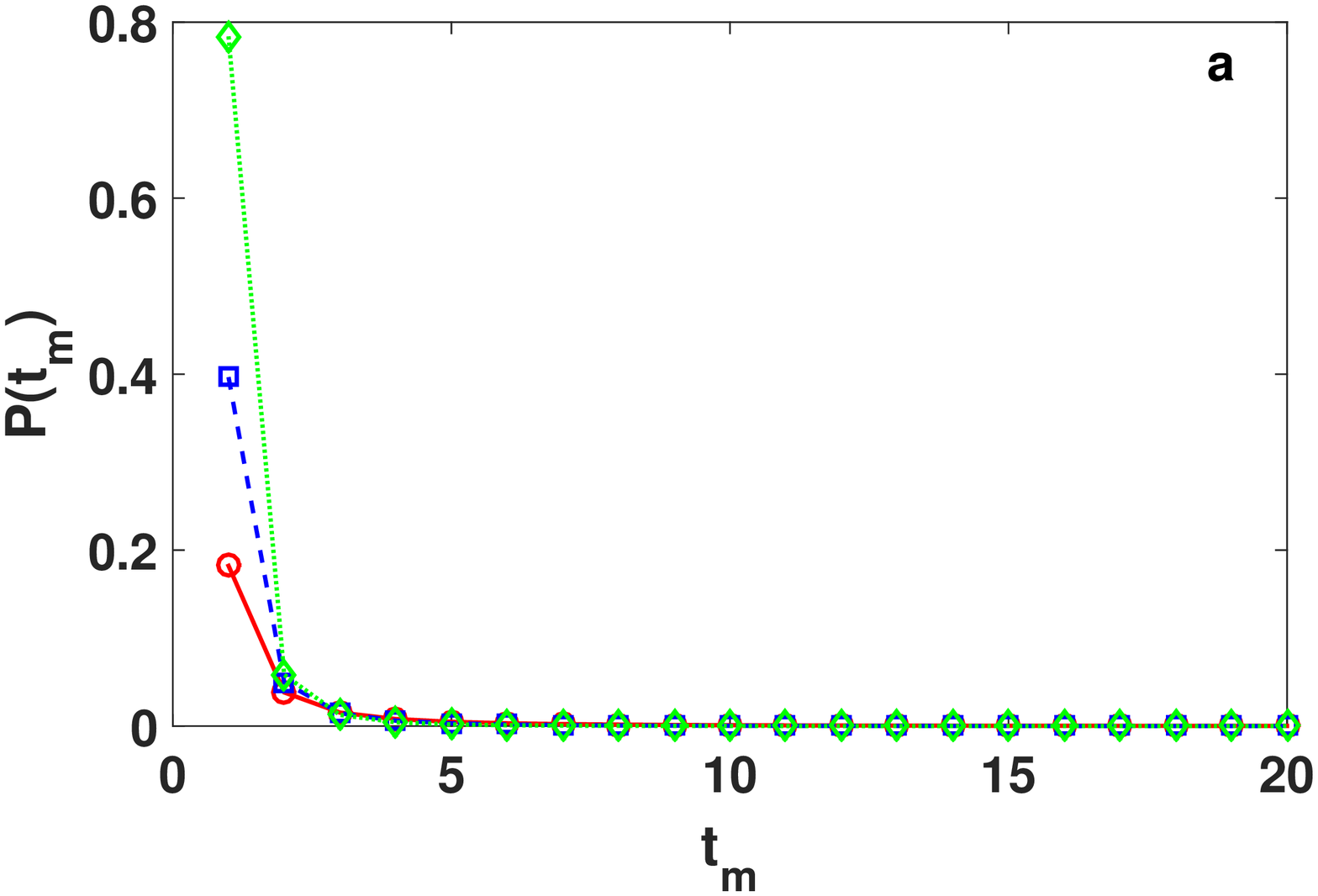}
 \caption{(color online) Plot of pdf $P(t_m)$ before first passage time $t_f$ for the small (a) $t_m$ as well as for the (b) large $t_m$ asymptotes with unbiased case, i.e., $\mu(t)=0$ and $D(t)=2kt^{\alpha}$. Here, we use $k=1.0$, $\alpha=0.5$ (red circle); $\alpha=1.0$ (blue square) and $\alpha=1.5$ (Green triangle).}
\label{fig:energy}
\end{center}
\end{figure*}
}
\newcommand{\figSix}
{
 \begin{figure*}[h]
 \begin{center}
 \includegraphics[width=3in,height=2.3in]{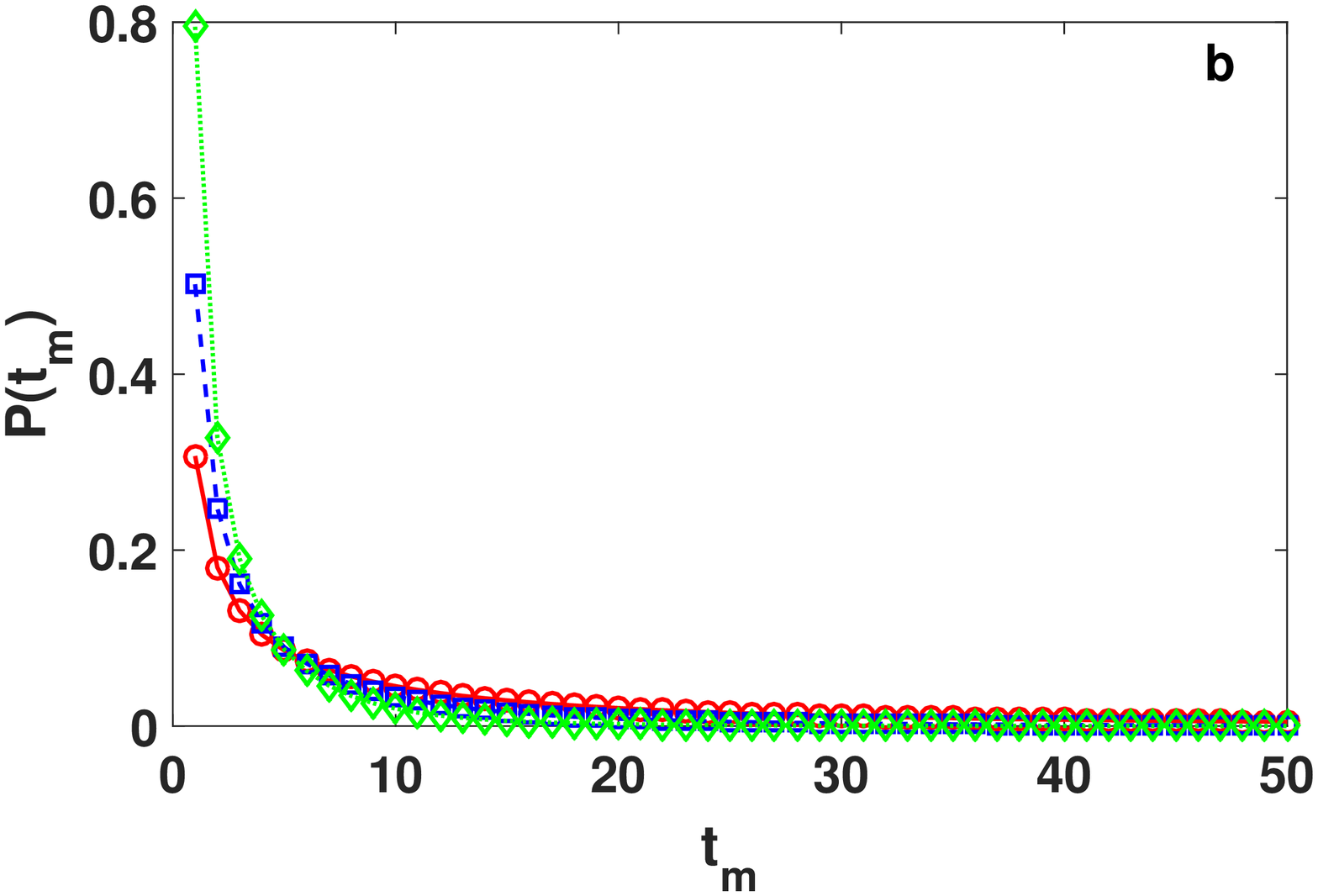}
 \includegraphics[width=3in,height=2.3in]{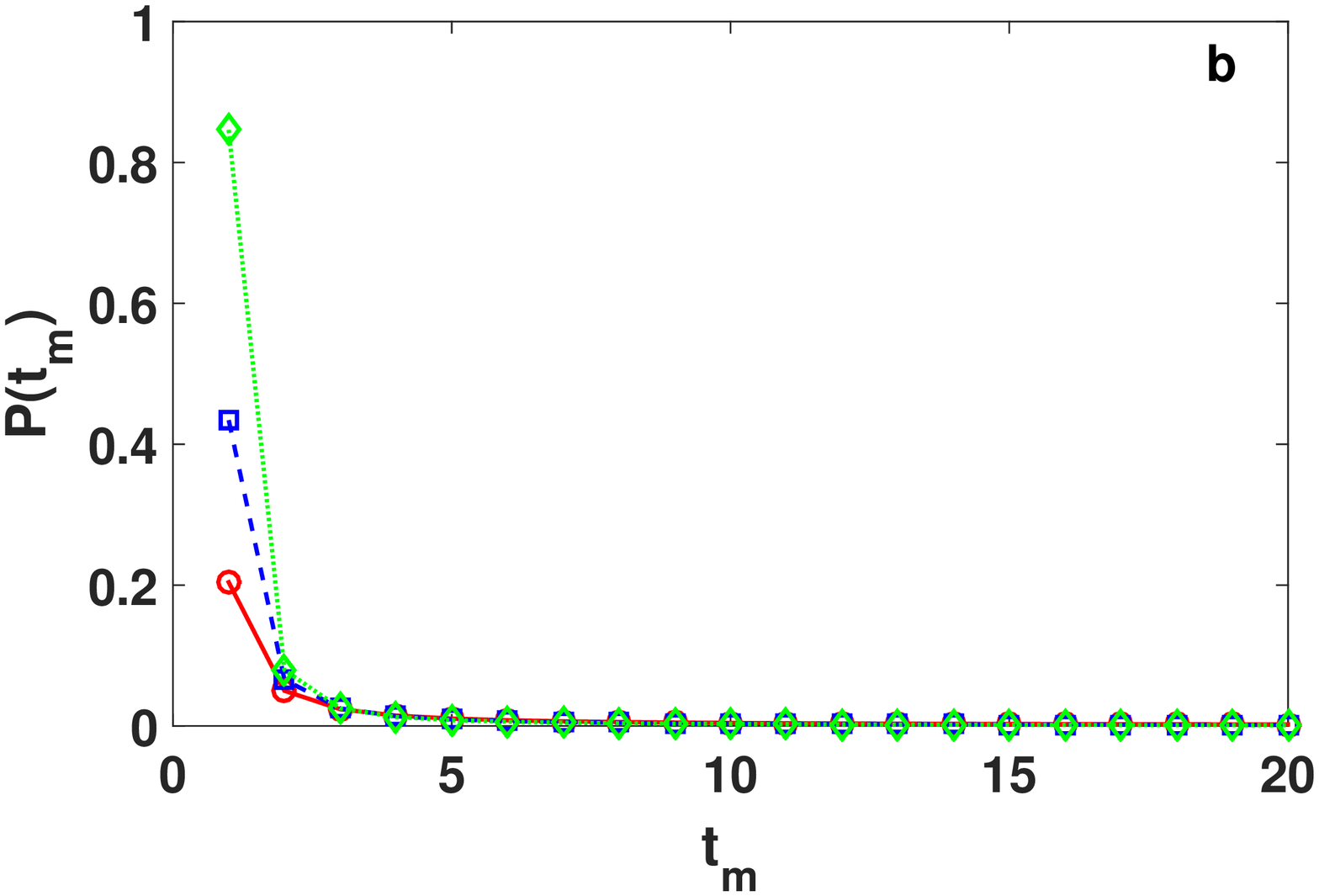}
 \caption{(color online)Plot of pdf $P(t_m)$ before first passage time $t_f$ for the small (a) $t_m$ as well as for the (b) large $t_m$ asymptotes with biased case, i.e., $\mu(t)=qkt^{\alpha}$ and $D(t)=2kt^{\alpha}$. Here, we use $q=0.1$, $k=1.0$, $\alpha=0.5$ (red circle); $\alpha=1.0$ (blue square) and $\alpha=1.5$ (Green triangle). }
\label{fig:energy}
\end{center}
\end{figure*}
}

\begin{document}
\title{Study of Brownian functionals in physically motivated model with purely time dependent drift and diffusion}
\author{Ashutosh Dubey$^1$, Malay Bandyopadhyay$^1$ and A. M. Jayannavar$^{2,3}$}
\affiliation{1. School of Basic Sciences, Indian Institue of Technology Bhubaneswar, Bhubaneswar, India 751007\\
2. Institute of Physics, Sachivalaya Marg, Sainik School PO, Bhubaneswar, India, 751005.\\
3.Homi Bhabha National Institute, Training School Complex, Anushakti Nagar,
Mumbai-400085, India}
\begin{abstract}
 In this paper, we investigate a Brownian motion (BM) with purely time dependent drift and difusion by suggesting and examining several Brownian functionals which characterize the lifetime and reactivity of such stochastic processes. We introduce several probability distribution functions
(PDFs) associated with such time dependent BMs. For instance, for a BM with initial starting point $x_0$, we derive analytical expressions for : (i) the PDF $P(t_f|x_0)$ of the first passage time $t_f$ which specify the lifetime of such stochastic process, (ii) the PDF $P(A|x_0)$ of the area A till the first
passage time and it provides us numerous valuable information about the effective reactivity of the process, (iii) the PDF $P(M)$ associated with the maximum size M of the BM process before the first passage time, and (iv)the joint PDF $P(M; t_m)$ of the maximum size M and its occurrence time
$t_m$ before the first passage time. These distributions are examined for the power law time time dependent drift and diffusion. A simple illustrative example for the stochastic model of water resources availability in snowmelt dominated regions with power law time dependent drift and diffusion and
 is demonstrated in details. We motivate our study with approximate calculation of an unsolved problem of Brownian functionals including inertia.
\end{abstract}
\pacs{05.40-a, 05.20.-y, 75.10.Hk}
\maketitle
\section{Introduction}
Brownian motion with purely time dependent drift and
diffusion are ubiquitous in geophysical, environmental
and biophysical processes. One can identify numerous
geophysical and environmental processes which occur under the crucial effect of external time dependent and random forcing, e.g., the change between the snow-storage and snow-melt phases \cite{a,b}, outbreak of water-borne
diseases \cite{c,d}, the life cycle of tidal communities \cite{e,f,g}, and many more. Stochastic models with time dependent drift and diffusion terms are extensively used in the study of neuroscience \cite{h,i,j}. One of the most useful tool to tackle such stochastic processes is the Fokker-
Planck formalism \cite{k,l}. In this formalism, different realizations of a system are narrated in terms of probability density which denotes the system in a given state at a certain instant and the theoretical description of such 1-D diffusion motion is governed by :
\begin{equation}
\frac{\partial P(x,t)}{\partial t}=-\mu(t)\frac{\partial P(x,t)}{\partial x}+D(t)\frac{\partial^2 P(x,t)}{\partial x^2}.
\end{equation}
In this respect several interesting questions of wide inter-
est can be raised such as, (i)the probability of finding the
system in a certain domain at a certain instant (survival
probability), (ii)the pdf of time $P(t_f|x_0)$ at which the
system exit a certain domain first time (known as first
passage time $t_f$ ) starting from initial point $x_0$, (iii)the
pdf $P(M)$ of the maximum value of a BM process be-
fore of its first passage time, and (iv)the joint probability distribution
$P(M; tm)$ of the maximum value M and its occurrence
time $t_m$ before the first passage time of the BM process.\\
\indent
All the above mentioned PDFs are calculated and discussed for simple Wiener and Ornstein-Uhlenbeck processes \cite{l,m,n} as well as in the context of DNA breathing dynamics \cite{malay}. But, all these discussions are
based on constant drift and diffusion terms. However, the
extension to a time dependent drift and time dependent
diffusion terms are not straightforward. This is mainly
because of the fact that the system has broken both the
space and time homogeneity. Several attempts are made
to study BM process with purely time dependent drift
and diffusion terms. One of the main work on BM with
time dependent drift is barrierless electronic reactions in
solutions \cite{amj1,amj2,amj3}. Generalizing the Oster-Nishijima model
\cite{onj} to the low viscosity limit or the inertial limit, the
authors observed a strong dependence on friction and
temperature of the decay rate even in the absence of the
barrier, which agrees well with numerical simulation of
the full Lanevin equation \cite{amj2}. A series of works on
stochastic resonance for time dependent sinusoidal drift
is analyzed in Refs. \cite{amj4,amj5,amj6,amj7}. The first passage time
statistics for a Wiener process with an exponential time
dependent drift term are analyzed in the context of neu-
ron dynamics in Refs. \cite{uph1,uph2}. Also, recent studies of DNA unzipping under periodic forcing need to be mentioned \cite{sanjay,alex,swan}. Recently, Molini {\it et. al} \cite{molini} make a study on BM with purely time dependent drift and diffusion terms.\\
\indent
In this work, we extend above mentioned works \cite{l,m,n,malay,molini} by incorporating several PDFs of Brownian motion i.e. $P(A|x_0)$, $P(M)$ and $P(M; t_m)$ for a BM with purely time dependent drift and diffusion terms. One of the main objective of this work is to incorporate inertial
effect in Brownian functional study and to our best of
knowledge, it is the first attempt to incorporate inertial
effect in first passage study which is one of the impor-
tant unsolved problem. The other objective of this work is to advertise for the use of the recently studied backward Fokker-Planck (BFP) method \cite{snm1} and the path decomposition (PD) method \cite{snm2}. Both the BFP and PD methods are based on the Feynman-Kac formalism \cite{kac}
and both of them are first time used for exploring BM
process with purely time dependent drift and diffusion
terms. Both the techniques are extensively used in study-
ing many aspects of classical Brownian motion, as well
as for exploring different problems in computer science
and astronomy \cite{snm1,snm3,snm4}. For the first time, we consider
these elegant methods to study the Brownian functionals
for a BM with purely time dependent drift and diffu-
sion. Unlike the standard FP treatment \cite{q,r,s} which
yields distribution functions directly, we derive and solve
differential equations for the Laplace transforms of var-
ious Brownian functionals in the BFP method. On the
other hand, we can utilize the PD method to calculate
the distribution functions of interest by splitting a rep-
resentative path of the dynamics into parts with their
appropriate weighage of each part separately. This fact
is justifiable by considering the Markovian property of
the dynamics.\\
\indent
The paper is organized as follows. In section II, we dis-
cuss our BM process model with purely time dependent
drift and diffusion terms. Then we discuss several dis-
tribution functions of interest and their relevances. The
BFP and PD methods are explained in short. In Sec.
III, we introduce several PDFs for a BM with power law
time dependent drift and diffusion terms. We illustrate the example
of fresh water availability in summer in the snowmelt dominated regime with the power law time dependent drift and diffusion terms. We conclude our paper in section IV.
\section{Model,Methods and Measures}
\subsection{Model}
We are interested with those kind of problem where time-dependent random forcing is predominant. Hence, the Fokker-Planck description of such problem can be made through Eq. (1). The associated stochastic differential equation for the state variable x(t) is given by :
\begin{equation}
dx(t)=\mu(t)dt+D(t)dW(t),
\end{equation}
where, $\mu(t)$ is the purely time dependent drift term, $D(t)$ denotes the diffusion term, and $W(t)$ is a Wiener process with Gaussian distribution. The Wiener process is an idealized statistical descriptions that
apply to many physical systems \cite{l,m,von}. One of the most elegant  theoretical method to tackle such kind of stochastic processes is
the Fokker-Planck (FP) formalism \cite{l,m,von}. In this formalism, one can describe  different realizations of a system by the
probability density. One can find the system in a given state at a
certain time, and the corresponding diffusion equation describes its temporal
evolution. Several interesting questions  related to such stochastic systems are of wide interest in several areas \cite{l,m,von}. One of the main interest in this field is to find the probability density for  the system remains in a certain domain at a given instant and the moment at which the system
escapes it for the first time. Due to the stochastic nature of the system, different realizations of the system leave a certain domain at different times and it is natural to consider the statistical properties of this random variable. Other interesting questions related with such first passage statistics are (i) finding the probability density $P(A|x_0)$ of area under a path (ii) probability density $P(M)$ of maximum size and (iii) the joint probability density $P(M,t_m)$ of maximum size and its occurrence time $t_m$.\\
\subsection{Methods}
In one dimension, the first passage statistics related problem are basically
formulated by considering a state variable which evolves stochastically
according to a given law in its phase space. We are mainly concerned about the instant when the variable leaves a certain domain for the first time. To deal with such problem a number of several methods or approaches had been described in REfs. \cite{l,m,snm1,von}. Here, we describe two elegant methods (i)Backward Fokker-Planck (BFP) method and (ii) Path decomposition method (PD).
\subsubsection{Backward Fokker-Planck Method (BFP)}
Following Ref. \cite{snm1},  we can introduce a general description to compute the PDF of a Brownian functional in a  time interval $\lbrack0,t_f\rbrack$, where $t_f$ is the first passage time of the process. Thus, one can introduce a functional to calculate different statistical properties of a Brownian functional :
\begin{equation}
T=\int_{0}^{t_f}U(x(\tau))d\tau,
\end{equation}
where, $x(\tau)$ is a Brownian path which follows differential Eq.(2) and it starts at $x_0$ at time $\tau=0$ and continues up to $\tau=t_f$. Here, $U(x(\tau))$ is a specified function of the path and its form depends on the quantity we are interested to calculate. For example, if we are interested to calculate first passage time one should choose $U(x(\tau))=1$. On the other hand, for the area distribution one should consider $U(x(\tau))=x(\tau)$. One can easily understand that $T$ is a random variable which can take different values for different Brownian paths. The main goal is to calculate probability distribution $P(T,t_f|x_0)$. Now, one may note that the random variable $T$ can be only positive for our choice of $U(x(\tau))$, Thus, one may consider the Laplace transform of the distribution $P(T|x_0)$ :
\begin{eqnarray}
Q(x_0,p)&=&\int_{0}^{\infty}dT P(T|x_0)\exp(-pT)\nonumber \\
&=& <\exp(-p\int_{0}^{t_f}U(x(\tau))d\tau)>.
\end{eqnarray}
Here, the angular bracket denotes the average over all possible paths starting at $x_0=0$ at $\tau=0$ and ending at the first time they cross the origin. For simplicity, we will drop the variable p in the function $Q(x_0,p)$ in the rest of our paper . Now, to derive a differential equation for $Q(x_0)$, we follow the method described in Ref. \cite{snm1}. Thus, we split the interval $\lbrack 0,t_f \rbrack $ into two parts. During the first interval
$\lbrack 0,\Delta\tau\rbrack$, the path starts from $x_0$ and propagates up to $x_0 + \Delta x$. In the second interval $\lbrack \Delta \tau,t_f \rbrack$, the path starts at $x_0 + \Delta x $ and ends at 0 at time $t_f$ . Here, $\Delta\tau$ is a fixed, infinitesimally small time interval. To leading order in $\Delta\tau$, we obtain :
$\int_{0}^{t_f}U(x(\tau ))d\tau \simeq U(x_0)\Delta\tau +\int_{\Delta\tau}^{t_f} U(x)d\tau$. As a result of that one can obtain from Eq. (4) :
\begin{eqnarray}
Q(x_0) &\simeq & \exp(-pU(x_0))< Q(x_0 + \Delta x)>_{\Delta x}\nonumber \\
&\simeq & (1-pU(x_0)\Delta\tau)<Q(x_0 + \Delta x)>_{\Delta x}.
\end{eqnarray}
Here, the angular bracket denotes the average over all possible realizations of $\Delta x$. Now, one can obtain from the dynamical equation for a free Langevin particle, i.e. from $\frac{dx}{dt}=\xi(t)$ that $\Delta x=\xi(0)\Delta \tau$. Now, expanding $Q(x_0 + \Delta x)$ in powers of $\Delta\tau$,
and taking the averages over the noise by using the facts  $<\xi(0)> = 0$ and $<\xi^2(0)> = 1/\Delta\tau$ as $\Delta\tau\rightarrow 0$, one obtains, to lowest order in $\Delta\tau$, the ordinary differential equation :
\begin{equation}
\frac{1}{2}\frac{d^2Q(x_0)}{dx_0^2}-pU(x_0)Q(x_0)=0.
\end{equation}
{\it Boundary Conditions:} Equation (6) is valid in the regime  $x_0\epsilon \lbrack 0,\infty\rbrack$ with the following boundary conditions : (i) As the initial position $x_0\rightarrow 0$, the first passage time vanishes which gives us  $Q(x_0=0)=1$, (ii)on the other hand, as $x_0\rightarrow \infty$, the first passage time diverges which results in $Q(x_0\rightarrow\infty)=0$.\\
\indent
Thus, our scheme will be as follows. We can solve the differential Eq. (6), termed as the BFP equation \cite{snm1}. By solving Eq.(6) with appropriate boundary condition as mentioned above provides us the Laplace transformed
pdfs of various quantities which are determined by the choice of U(x). Now, inverting the Laplace transform with respect to p, one can obtain the desired pdf $P(T|x_0)$. On the other hand, the standard Fokker-Planck
method adopted in Refs. \cite{l,m} yields the distribution
function P(x,t )directly. Thus, these two approaches are distinct, providing complementary information.
\subsubsection{The path decomposition method (PD)}
The basic principle of this PD method is very simple. Since, our
motion in Eq.(2) is Markovian one can break a typical path
into two parts. Thus, the weightage of the whole path is the
product of the weights of the two split parts \cite{snm1}.
Thus, the joint probability distribution $P(M,t_m)$ of the maximum bubble size M and the occurrence time $t_m$ at which this maximum occurs before first passage. Now, integrating over M, one can obtain the marginal distribution $P(t_m)$. The basic process to compute $P(M,t_m)$ by splitting a typical
path into two parts, before and after $t_m$. Here, weights $W_L$ and $W_R$ are the weighage of the path before and after $t_m$. As a matter of fact, the total weight W of the whole path is :
\begin{equation}
W = W_L \times W_R
\end{equation}
On the left-hand side of $t_m$, the path propagates from $x_0$ at
$t = 0$ to $M − \epsilon$ at $t = t_m$, without attaining the value
$0$ or $M$ during the interval $\lbrack0,t_m\rbrack$ \cite{snm2}. Now, The weight $W_L$ can be determined by using a path-integral treatment based on
the Feynman-Kac formalism. Let us we denote $q(x_0)$ be the probability that the motion described by Eq. (2) exits the interval $\lbrack 0,M \rbrack$ for the first time through the origin. Thus, $q(x_0)$ is the cumulative probability that the maximum before the first-passage time is $\leqslant M$. It is known that this function satisfies two boundary conditions: (i) $q(0) = 1$ and
(ii) $q(M) = 0$. Let us consider a function $\phi_{\Delta\tau}(\Delta x)$ which gives us the distribution function of a small displacement $\Delta x$ in time $\Delta \tau\rightarrow 0$. Now, using the Markovian property of the dynamics (2), one can show that :
\begin{equation}
q(x_0)=\int q(x_0+\Delta x)\phi_{\Delta\tau}(\Delta x)d(\Delta x),
\end{equation}
Now, making a Taylor expansion of $q(x_0+\Delta x)$ and averaging over $\Delta x= \xi(0)\Delta \tau$, and using $<\xi(0)>=0$ and $<\xi^2(0)> = 1/\Delta\tau$. Thus, to the leading order in $\Delta \tau$ we obtain
\begin{equation}
\frac{d^2q(x_0)}{dx_0^2}=0.
\end{equation}
Now, solving the above equation with the help of above mentioned boundary boundary condition, one can obtain :
\begin{equation}
q(x_0)=1-\frac{x_0}{M}.
\end{equation}
Now, differentiating $q(x_0)$ with respect to $M$ we obtain :
\begin{equation}
P(M)=\frac{x_0}{M^2}
\end{equation}
Now, the $W_R$ is obtained as
\begin{equation}
W_R=q(M-\epsilon)=\frac{\epsilon}{M}.
\end{equation}
On the other hand, the weight $W_L$ can be obtained from the white noise is Gaussian and the probability of a path is given by :
\begin{equation}
P\lbrack \lbrace x(\tau)\rbrace\rbrack \propto \exp \Big\lbrack -\frac{1}{2}\int_{0}^{t}d\tau \Big(\frac{dx}{d\tau}\Big)^2\Big\rbrack.
\end{equation}
Then, the weight $W_L$ is then obtained as a sum over contributions from all possible paths :
\begin{eqnarray}
W_L &\propto & \int_{x(0)=x_0}^{x(t_m)=M-\epsilon}{\mathcal{D}}x(\tau)\exp \Big\lbrack -\frac{1}{2}\int_{0}^{t}d\tau \Big(\frac{dx}{d\tau}\Big)^2\Big\rbrack \nonumber \\
&\times & \prod_{\tau=0}^{t_m}\theta\lbrack x(\tau)\rbrack \prod_{\tau=0}^{t_m}\theta\lbrack M- x(\tau)\rbrack,
\end{eqnarray}
where, $\prod_{\tau=0}^{t_m}\theta\lbrack x(\tau)\rbrack$ and $\prod_{\tau=0}^{t_m}\theta\lbrack M-x(\tau)\rbrack$ enforce the requirements that the path does not cross either the level $0$ or the level $M$ for times between $0$ and $t_m$. Now, following Feynman-Kac \cite{kac}, the path integral can be identified with the propagator $<M-\epsilon|e^{-\hat{H}t_m}|x_0>$, corresponding to the quantum Hamiltonian $\hat{H}$ of a single particle of unit mass,
\begin{equation}
\hat{H}=-\frac{1}{2}\frac{d^2}{dx^2}+V(x),
\end{equation}
with potential energy $V(x)=0$ for $0<x<M$ and $V(x)=\infty$ for $x=0$ and $x=M$. Note, that the infinite potential energy at $x=0$ and at $x=M$ enforces the requirement that the path never crosses either the level 0 or level M. Finally,
\begin{equation}
W_L=\Big(\frac{x_0}{M-\epsilon}\sum_{n=1}^{\infty}e^{-E_nt_m}\psi_n(M-\epsilon)\psi_n(x_0),
\end{equation}
where, $\psi_n(x)$ and $E_n$ are the eigenfunctions and eigenenergies, respectively for the Hamiltonian $\hat{H}$.
\subsection{Measures}
Our primary focus is on several first-passage Brownian
functionals of physical relevance. We consider the following quantities and explore their pdfs. In this context, we explore a physical phenomenon of snowmelt dynamics for the fresh water availability in summer.\\
(i){\it First passage time or lifetime of the stochastic process}: The first-passage time pdf $P(t_f |x_0)$, i.e., the pdf of the time of touching the origin first time with initial size $x_0$, provides the information about the lifetime of the stochastic process. A related quantity is the survival probability $C(x_0,t ) = 1 - \int_{0}^{t_f} P(t_f|x_0 )dt_f$ of the process. This survival probability is an experimentally measurable quantity. For example, in the context of DNA breathing dynamics $C(x_0,t )$ can be inferred from experiments by measuring fluorescence correlations of a tagged DNA \cite{bonnet1,bonnet2}. In the snow melt dynamics, our key stochastic variable is the total potential water availability, H (in terms of water equivalent from both snow and rainfall). Thus, the survival probability $C(H_0,t)$ for a given initial snow water equivalent $H_0$ and the pdf of first passage time $P(t_f|H_0)$ are very much useful quantities to offer important information about the timing between melting of snow and fresh water availability in summer under different climatic scenarios. \\
(ii){\it Area under a path:} If we consider a typical path which is described by Eq. (2), one can define the area under such a path before the first-passage time as $A =\int_{0}^{t_f} x(t)dt$. The interesting quantity is its pdf
$P(A|x0)$ with an initial value $x_0$. This quantity is of interest because it provides a measure for the effectiveness of the corresponding stochastic processes. For example, if we consider the snow melt process, then $P(A|H_0)$ gives us the information about the average total snow water equivalent with initial value $H_0$. While the first-passage time distribution provides information about the lifetime, it does not contain any hint of the average total water equivalent before full melting. Quantities (i), (ii)can be calculated below by following the BFP method discussed in Sec. IIB-1.\\
{\it Maximum size M:} The other proposed measure
for quantifying reactivity of the process is the distribution of the
maximum  size before the first-passage time, P(M). Let us
consider again snow melt process. Now, the pdf $P(M)$ provide us about the information about the maximum total available fresh water equivalent before total melting of snow.\\
{\it Maximum size M and the corresponding time tm:}
The joint probability distribution function $P(M,t_m)$ can be
investigated here by following the PD method, which is based
on the Feynman-Kac formalism \cite{kac} (see Sec. IIB-2). Using
this pdf, one can further calculate the distribution function
$P(t_m)$ of the time at which the process attains its maximum
size before hitting the origin. This latter pdf is of interest because it
provides information about the (average) time of occurrence
of the biggest size before hitting the origin.
\section{Snowmelt dynamics}
Snowmelt is one of the main source of freshwater for many regions of the world and the snowmelt process is very much sensitive to temperature and precipitation fluctuations \cite{barnett1,barnett2}. Snow dynamics is basically consists of two phases : (a) an accumulation phase in which snow water equivalent (i.e. the amount of liquid water available by total and instantaneous melting of the entire snowmass) rises to its seasonal maximum $Q_0$ and the other one is (b)the depletion phase where the whole snowpack gradually decreases (release of stored water content) due to temperature fluctuation. To describe such a complex dynamics one needs a lot of physical parameters. Now, we are trying to build a simplified stochastic model which can describe the total water equivalent from both snow and rainfall during the melting season, as driven by both precipitation (solid to liquid transition) and increasing air temperature.
Due to simplification of the stochastic model, we consider the total potential water availability (in terms of water equivalent) as the main stochastic variable. Here, we neglect any other effects connected with snow percolation and metamorphism etc. \cite{rango}. The predominant factors which govern the fresh water availability in the warm season are increasing air temperature and liquid precipitation. Accordingly, we assume the melting phase can be described by a power-law time dependent drift directed towards the total melting of the snowpack. On the other hand, positive and negative exponents of power-law diffusion usually represent precipitation events and pure melting periods, respectively. Following the "degree-day" approach with time-varying melting-rate coefficients, one can assume the melting process can be described by a linear function of time \cite{rango}. Considering a power-law form for drift and diffusion during the melting season, the dynamics of the total water equivalent from both snow melting and precipitation at a given point in space can be reasonably described by the Langevin equation \cite{bras} :
\begin{equation}
dH=-\mu(t)dt + D(t)dW(t),
\end{equation}
where, the drift part $\mu(t)=kt^{\alpha}$ represents the accumulation or depletion with a rate constant $k$ and the diffusion rate is given by $D(t)=\sqrt{2kt^{\alpha}}$. Also, both the rainfall and snowmelt contributions are included in $Q$. Here, we assume that the drift and the diffusion follow the same power law with exponent $\alpha$. This is a reasonable assumption in the sense that the snow melt is most predominant in the summer time i.e. the process is expected to increase its variability during warm season \cite{molini}. The initial value of the snow water equivalent (SWE), $H_0$, is the accumulated snow during the cold season.\\
\indent
The Fokker-Planck equation corresponding to the differential Eq. (17)
\begin{equation}
\dfrac{\partial p(H,t|H_{0})}{\partial t}=-\mu(t)\dfrac{\partial p(H,t|H_{0})}{\partial H}+ D(t)\dfrac{\partial^{2}p(H,t|H_{0})}{\partial H^{2}}.
\end{equation}
Now, we can use the following transformation equations to go from $(H,t)$ to $(z,\tau)$ space
\begin{equation}
\tau=\int D(t) dt + B,
\end{equation}
and
\begin{equation}
z=H+\int\mu(t)dt +C.
\end{equation}
Using the above mentioned transformation equations one can reduce Eq. (18) into a constant co-efficient free diffusion equation form :
\begin{equation}
\frac{\partial p(z,\tau)}{\partial \tau}=\frac{\partial^2p(z,\tau)}{\partial z^2},
\end{equation}
\figOne
\subsection{PDF of first Passage Time : $P(t_f|H_0)$}
Using the backward Fokker-plank method one can obtain the BFP equation
\begin{equation}
\dfrac{1}{2}\dfrac{d^{2}Q(z_{0},\tau)}{d z_{0}^{2}} -pU(z_{0})Q(z_{0}) = 0.
\end{equation}
Substituting $U(z_{0})= 1$ in equation (22), we obtain
\begin{equation}
\dfrac{1}{2}\dfrac{d^{2}Q}{d z_{0}^{2}} -pQ(z_{0}) = 0.
\end{equation}
The general solution of equation (23) is
\begin{equation}
Q(z_{0})=e^{-\sqrt{2p} z_{0}}
\end{equation}
Inverting the Laplace transform with respect to p gives the pdf of the first passage time for $\tau_f$
\begin{equation} P(\tau_{f}|z_{0})=\dfrac{z_{0}}{\sqrt{2\pi}}\dfrac{e^{-z_{0}^{2}/2\tau_{f}}}{\tau_{f}^{3/2}}
\end{equation}
Again transforming above equation into original variables $H$ and $t$ by using equations (19) and (20), we get
\begin{eqnarray}
P(t_{f}|H_{0})&=&D(t_f)\dfrac{1}{\sqrt{4\pi}}\dfrac{[H_{0}+\int_{0}^{t_{f}}\mu(t)dt]}{[1/2\int_{0}^{t_{f}}\sigma^{2}(t)dt]^{3/2}}\nonumber \\ &\times&\exp\bigg[-\dfrac{(H_{0}+\int_{0}^{t_{f}}\mu(t)dt)^2}{2\int_{0}^{t_{f}}\sigma^{2}(t)dt} \bigg]
\end{eqnarray}
Let us consider two different cases for the time dependent drift and diffusion :	(1)$D(t)=\sqrt{2kt^{\alpha}}$, $\alpha>-1$, and $\mu(t)=0$.\\
Now, substituting $\mu(t)$ and $D(t)$ in equation (26), we obtain
\begin{equation}
P(t_{f}|H_{0})=D(t_f)\dfrac{1}{\sqrt{4\pi}}\dfrac{H_{0}}{\Big[k\dfrac{t_f^{\alpha+1}}{\alpha+1}\Big]^{3/2}} \exp \Big[-\dfrac{H_{0}^2}{2k\dfrac{t_f^{\alpha+1}}{\alpha+1}}\Big]
\end{equation}
(2)Case 2 :  proportional power-law diffusion and drift i.e. $\mu(t)=qkt^{\alpha}$ and $D(t)=\sqrt{2kt^{\alpha}}$; then the first passage time distribution is given by
\begin{eqnarray}
&&P(t_{f}|H_{0})= \dfrac{\Big(H_{0}+\dfrac{qkt_{f}^{\alpha+1}}{\alpha+1}\Big)(1+\alpha)^{3/2}}{2\sqrt{\pi A}t_{f}^{(3+\alpha)/2}} \nonumber \\ &\times& \exp\Big[-\dfrac{t_{f}^{-(\alpha+1)}(kqt_{f}^{\alpha+1}+H_{0}+\alpha H_{0})^2}{2k(\alpha+1)}\Big]
\end{eqnarray}
\subsection{PDF of area till the first passage time: $P(A|H_0)$}
Whereas the $P(t_f|H_0)$ can supply the important information about the time of melting and summer fresh water availability, the pdf $P(A|H_0)$ will supply us the useful information about the total summer fresh water availability under different climatic conditions.\\
We can compute the distribution of $A$,i.e., $P(A|H_0)$ by substituting $U(z_{0})=z_{0}$ in equation (13):
\begin{equation}
\dfrac{d^{2}Q}{d z_{0}^{2}} -2p z_{0}Q(z_{0}) = 0
\end{equation}
The general solution of equation (29) is
\begin{equation}
Q(z_{0})=A_{1} Ai(2^{1/3}p^{1/3}z_{0})+B_{1}Bi(2^{1/3}p^{1/3}z_{0}),
\end{equation}
where $Ai(z)$ is the Airy function. Now, applying the boundary conditions :\\  1.$ Q(z_{0})=0~~~~$ when $z_{0}\rightarrow \infty$\\
2.$ Q(z_{0})=1~~~~$ when $z_{0}\rightarrow 0$, we obtain
\begin{equation}
Q(z_{0})=3^{2/3}\Gamma(2/3)Ai(2^{1/3}p^{1/3}z_{0})
\end{equation}
Taking the inverse Laplace transform
\begin{equation} P(A(\tau)|z_{0})=\dfrac{2^{1/3}}{3^{2/3}\Gamma(1/3)}\dfrac{z_{0}}{A(\tau)^{4/3}}\exp\Big[-\dfrac{2z_{0}^3}{9A(\tau)}\Big]
\end{equation}
Again transforming above equation into original variables $A(t)$ and $t$ by using equation (19) and (20), we obtain
\begin{eqnarray} P[A(t)|H_{0}]&=&D(t)\dfrac{2^{1/3}}{3^{2/3}\Gamma(1/3)}\dfrac{H_{0}+\int_{0}^{t_{f}}\mu(t)dt}{[A(t)]^{4/3}}\nonumber \\
&&\exp \Big[-\dfrac{2[H_{0}+\int_{0}^{t_{f}}\mu(t)dt]^3}{9A(t)}\Big]
\end{eqnarray}
{\it Case (1): Unbiased power law time dependent diffusion }\\
In this case one can consider $D(t)=\sqrt{2kt^{\alpha}}$, $\alpha>-1$, and $\mu(t)=0$. Now, substituting the above mentioned values of $D(t)$ and $\mu(t)$ in Eq. ()
we obtain the pdf of area till $t_f$ :
\begin{eqnarray} P(A|H_{0})&=&(kt_{f}^{\alpha})\dfrac{2^{1/3}}{3^{2/3}\Gamma(1/3)}\dfrac{H_{0}}{[A(t)]^{4/3}}\nonumber \\
&& \times \exp\Big[-\dfrac{2H_{0}^3}{9A(t)}\Big]
\end{eqnarray}
\figThree
{\it (2)Case 2 :  Proportional power-law diffusion and drift}\\
 In the case of proportional power-law diffusion and drift, one may consider $\mu(t)=qkt^{\alpha}$ and $D(t)=\sqrt{2kt^{\alpha}}$, the PDF of area till the first-passage time is given by:
\begin{eqnarray} P(A|H_{0})&=&\dfrac{2^{1/3}}{3^{2/3}\Gamma(1/3)}\dfrac{(kt_{f}^{\alpha})[(\alpha +1)H_{0}+qkt_{f}^{\alpha+1}]}{(\alpha+1)[A(t)]^{4/3}}\nonumber \\
&&\exp\Big[-\dfrac{2(\alpha H_{0}+H_{0}+qkt_{f}^{\alpha+1})^{3}}{9(\alpha+1)^3A(t)}\Big]
\end{eqnarray}
\subsection{Joint probability distribution of maximum and its occurrence before first passage time : $P(M,t_m)$}
The joint probability distribution of maximum and its occurrence before first passage time,$P(M,t_m)$, provides important information about the maximum available fresh water equivalent in summer as well as the exact timing of it. In that sense it is one of the important quantity to study. Now, following the Path decomposition method discussed in Section IIB-2 as well as in Ref. \cite{snm2}, we can obtain the exact expressions of joint probability distribution $P(M,t_m)$ for the two cases of power law.\\
{\it Case (1)Unbiased power law time dependent diffusion}\\
In this case one can consider $D(t)=2kt^{\alpha}$, $\alpha>-1$, and $\mu(t)=0$.
Thus, the joint probability distribution $P(M,t_m)$ is given by
\begin{eqnarray}
P(M,t_{m})&= &(kt^{\alpha})\dfrac{\pi}{M^3} \sum_{n=1}^{\infty}(-1)^{n+1}n \sin\Big(\dfrac{n\pi x_{0}}{M}\bigg)\nonumber \\
 &&\exp\Big(-\dfrac{n^2 \pi^2}{2M^2}k\dfrac{t_{m}^{\alpha+1}}{\alpha+1}\Big)
\end{eqnarray}
{\it (2)Case 2 :  Proportional power-law diffusion and drift}\\
 In the case of proportional power-law diffusion and drift, one may consider $\mu(t)=qkt^{\alpha}$ and $D(t)=\sqrt{2kt^{\alpha}}$, the joint probability distribution $P(M,t_m)$ is given by :
\begin{eqnarray}
P(M,t_{m})&= &(kt^{\alpha})\dfrac{\pi}{M^3} \sum_{n=1}^{\infty}(-1)^{n+1}n \sin\Big(\dfrac{n\pi }{M}(x_{0}+\dfrac{qkt_{m}^{\alpha+1}}{\alpha+1})\bigg)\nonumber \\
&& \exp\Big(-\dfrac{n^2 \pi^2}{2M^2}A\dfrac{t_{m}^{\alpha+1}}{\alpha+1}\Big)
\end{eqnarray}
It is very difficult to plot the joint probability distribution. So, we are interested on the marginal distribution $P(t_m)$.
\subsection{Marginal Distribution : $P(t_m)$}
The marginal distribution is given by
\begin{equation}
P(\tau_{m})=\int_{z_{0}}^{\infty}dM P(M,\tau_{m})
\end{equation}
Now, putting the expression of $P(M,\tau_m)$, one can obtain
\begin{eqnarray}
&&P(\tau_{m})	= \int_{z_{0}}^{\infty}dM\dfrac{\pi}{M^3} \sum_{n=1}^{\infty}(-1)^{n+1}n \sin\Big(\dfrac{n\pi z_{0}}{M}\Big)\nonumber \\
&&\times \exp\Big(-\dfrac{n^2 \pi^2}{2M^2}\tau_{m}\Big)\nonumber \\
&&= \sum_{n=1}^{\infty}(-1)^{n+1}n \pi\int_{z_{0}}^{\infty}\dfrac{dM}{M^3}  \sin\Big(\dfrac{n\pi z_{0}}{M}\Big) \exp\Big(-\dfrac{n^2 \pi^2}{2M^2}\tau_{m}\Big). \nonumber \\
\end{eqnarray}
Now, putting $u=\dfrac{n\pi z_{0}}{M}$ $\Rightarrow ~du=-\dfrac{n\pi z_{0}}{M^2}dM$, one can show that
\begin{equation}
P(\tau_{m})=\dfrac{1}{\pi \tau_{m}}\sum_{n=1}^{\infty}\dfrac{(-1)^{n+1}}{n}\int_{0}^{n\pi}du~ cos(u)\exp\bigg(-\dfrac{u^2}{2z_{0}^2}\tau_{m}\bigg)
\end{equation}
{\it Case I :Large $\tau_{m}$ asymptote ($\tau_{m}\gg z_{0}^2$)}\\
Introducing the variable $k=\sqrt{\dfrac{\tau_{m}}{2z_{0}^2}}$ in above equation, we obtain 
\begin{equation}
P(\tau_{m})=\dfrac{z_{0}~ log2}{(\tau_{m}^{3/2})}\dfrac{1}{\sqrt{2\pi}}
\end{equation}
Now, again transforming into (x,t) space we obtain 
\begin{equation}
P(t_{m})=D(t_m)\dfrac{log2}{\sqrt{2\pi}}\dfrac{[x_{0}+\int_{0}^{t_{m}}\mu(t)dt]}{[1/2\int_{0}^{t_{m}}D(t)dt]^{3/2}}
\end{equation}
{\it a. unbiased diffusion and $D(t)=2kt^\alpha$}\\
In this case, we obtain
\begin{equation}
P(t_{m})=\dfrac{log2}{\sqrt{2\pi k}}\dfrac{x_{0}(\alpha+1)^{3/2}}{t_{m}^{(\alpha+3)/2}}
\end{equation}
{\it Proportional power law drift and diffusion :$\mu(t)=qkt^\alpha$ and $D(t)=2kt^\alpha$}\\
In this case, the marginal distribution is given by
\begin{equation} P(t_{m})=\dfrac{log2}{\sqrt{2\pi}}\dfrac{(\alpha+1)^{1/2}}{k^{1/2}}\dfrac{[(\alpha+1)x_{0}+qkt_{m}^{\alpha+1}]}{(\alpha+1)(t_{m}^{\alpha+1})^{3/2}}
\end{equation}
{\it Case II : Small-$t_{m}$asymptote:} \\
In this limit $\tau_{m}\ll z_{0}^2$. Now, taking the Laplace transform of the Eq. ()
\begin{equation} \int_{0}^{\infty}d\tau_{m}~e^{-s\tau_{m}}P(M,\tau_{m})=\dfrac{sinh(z_{0}\sqrt{2s})}{M~sinh(M\sqrt{2s})}
\end{equation}
Let us consider s becomes much larger than $z_{0}^{-2}$ and $M^{-2}$, we get
\begin{equation}
\int_{0}^{\infty}d\tau_{m}~e^{-s\tau_{m}}P(M,\tau_{m})\approx \dfrac{e^{-\sqrt{2s}(M-z_{0})}}{M}	
\end{equation}
\figFive
\figSix
Taking the inverse Laplace transform
\begin{equation}
P(M,\tau_{m})\approx \dfrac{\tau_{m}^{-3/2}}{\sqrt{2\pi}}\dfrac{(M-z_{0})}{M}e^{\dfrac{(M-z_{0})^2}{2\tau_{m}}}
\end{equation}
Integrating the above equation over M in the limit $\tau_{m}\ll z_{0}^2$, we get
\begin{equation}
P(\tau_{m})\approx \dfrac{1}{z_{0}\sqrt{2\pi \tau_{m}}}
\end{equation}
Again,transforming into (x,t) variables we obtain
\begin{equation} P(t_{m})=D(t_m)\dfrac{1}{\sqrt{2\pi}}\dfrac{1}{[x_{0}+\int_{0}^{t_{m}}\mu(t)dt]}\dfrac{1}{\bigg[ \dfrac{1}{2}\int_{0}^{t_{m}}D(t)dt \bigg]^{1/2}}
\end{equation}
{\it Unbiased power law diffusion and $D(t)=2kt^\alpha$}\\
\begin{equation} P(t_{m})=\bigg(\dfrac{k(\alpha+1)}{2\pi}\bigg)^{1/2}\dfrac{1}{x_{0}}\dfrac{t_{m}^{\alpha}}{(t_{m}^{\alpha+1})^{1/2}}
\end{equation}
{\it Proportional power law time dependent drift and diffusion}\\
In this case $\mu(t)=qkt^\alpha$ and $D(t)=2kt^\alpha$
\begin{equation} P(t_{m})=(\alpha+1)^{3/2}\sqrt{\dfrac{k}{2\pi}}\dfrac{t_{m}^{\alpha}}{[(\alpha+1)x_{0}+qkt_{m}^{\alpha+1}]}\dfrac{1}{(t_{m}^{\alpha+1})^{1/2}}
\end{equation}
\section{Conclusions}
In this work, we analyze several relevant probability
distribution functions of various Brownian functionals
associated with the stochastic model for the total fresh water availability in mountain region incorporating both the temperature effect, snow accumulation and precipitation in the form of power law dependent drift $\mu(t)\sim qt^{\alpha}$ and diffusion constant $D(t)\sim kt^{\alpha}$.  Based on the backward Fokker-Planck method discussed in Ref.\cite{snm1}, we derive (i) the first-
passage time distribution $P(t_f|x0)$, providing informa-
tion about the lifetime of the stochastic process, (ii) the
distribution $P(A|x_0)$, of the area A covered by the ran-
dom walk till the first-passage time, measuring the re-
activity of stochastic processes, and (iii) the distribution
P(M), of the maximum size M before first passage time,
(iv) the joint probability distribution $P(M; t_m)$ of the
maximum size M and the time $t_m$ of its occurrence be-
fore the first passage time was also obtained by employing
the Feynman-Kac path integral formulation. The advan-
tage of the elegant methods adopted here is that they
produce results on various functionals by making proper
choices of a single term in a parent differential equation
with appropriate boundary condition. We are at present
studying these functionals for Brownian particle with in-
ertia. If we assume initial velocity to be zero the problem
is easily tractable. However, if we consider a Gibbsian
distribution of the initial velocity the problem is really
challenging and the work is under progress along this line
\cite{malay1}.\\
Also, this study is helpful in analyzing the effect of periodic forcing in DNA unzipping \cite{sanjay} or the study on the effect of terahertz field on DNA breathing dynamics \cite{alex,swan}.In the context of integrate-fire model with sinusoidal modulation of neu-
ron dynamics, the membrane voltage, V(t), is the stochastic variable under sinusoidal stimulus. In this context, $P(t_f|V_0)$ and $C(V_0,t)$ will provide important information about the timing of firing of neuron after reaching the threshold voltage $V_{th}$ with an initial value $V_0$ \cite{malay2}
\begin{acknowledgments}
MB acknowledge the financial support of IIT Bhubaneswar through seed money project SP0045. AMJ thanks DST, India  for award of J C Bose national
fellowship.
\end{acknowledgments}

\end{document}